\documentclass[11pt]{article}   
\pdfoutput=1

\usepackage{amsthm,amsmath,amssymb,bbold,xcolor,sectsty,latexsym,mathtools,hyperref,mathrsfs,slashed,bm,epsfig,enumerate, subcaption}
\usepackage[super]{nth}

\usepackage[textwidth=6.5in,top=1.2in,bottom=1.2in]{geometry}

\newcommand{\beq}{\begin{equation}}
\newcommand{\eeq}{\end{equation}}
\newcommand{\bea}{\begin{eqnarray}}
\newcommand{\eea}{\end{eqnarray}}

\newcommand{\Lagr}{\mathcal{L}}

\begin{document}

\thispagestyle{empty}
\begin{titlepage}
\nopagebreak

\title{ \begin{center}\bf Axions Mediated by a  Yukawa-like \nth{5} Force Boson \end{center} }
\title{ \begin{center}\bf Effects of New Forces on Scalar Dark Matter Solitons  \end{center} }

\vfill
\author{Alize Sucsuzer$^{a}$\footnote{alize.sucsuzer@tufts.edu}, \,\,\, Mark P.~Hertzberg$^{a}$\footnote{mark.hertzberg@tufts.edu}
\,\,\, Michiru Uwabo-Niibo$^{b}$\footnote{michiru@ibs.re.kr}}

\date{}

\maketitle

\begin{center}
	\vspace{-0.7cm}
	{\it  $^{a}$Institute of Cosmology, Department of Physics and Astronomy}\\
	{\it  Tufts University, Medford, MA 02155, USA\\
    \vspace{0.5cm}
    $^{b}$Cosmology, Gravity, and Astroparticle Physics Group, \\Center for Theoretical Physics of the Universe,\\ 
Institute for Basic Science (IBS), Daejeon}
	
\end{center}

\bigskip

\begin{abstract}
New long range forces acting on ordinary matter are highly constrained. However it is possible such forces act on dark matter, as it is less constrained observationally. In this work, we consider dark matter to be made of light bosons, such as axions. We introduce a mediator that communicates a new force between  dark matter particles, in addition to gravity. 
The mediator is taken to be light, but not massless, so that it can affect small scale galactic behavior, but not current cosmological behavior. As a concrete application of this idea, we analyze the effects on scalar dark matter solitons bound by gravitation, i.e., boson stars, which have been claimed to potentially provide  cores of galaxies. We numerically determine the soliton's profiles in the presence of this new force. We also extend the analysis to multiple mediators.
We show that this new force alters the relation between core density and core radius in a way that can provide improvement in fitting data to observed galactic cores, but for couplings of order the gravitational strength, the improvement is only modest. 
\end{abstract}

\end{titlepage}

\setcounter{page}{2}

\tableofcontents

\vspace{1cm}


\section{Introduction}

The nature of dark matter remains a fundamental question in modern physics. Observations are currently compatible with the dark matter interacting only gravitationally. However, it is possible that dark matter has other non-gravitational interactions. Self-interactions in the dark matter could lead to a kind of dynamical friction as galaxies pass through each other \cite{Spergel:1999mh, Kahlhoefer:2013dca} (for a review, see Ref.~\cite{Tulin:2017ara}). This is constrained by bullet cluster type observations to be $\sigma/m\lesssim \rm{cm}^2/\rm{g}$ \cite{Markevitch:2003at}. However, beyond that, the constraints are relatively weak. This leaves open the possibility of long (or moderately long) range forces between dark matter particles with ranges comparable to or smaller than galactic scales. If such new forces are infinitely long ranged, then one can anticipate new effects on cosmological scales. Since the evidence for standard cold dark matter on cosmological scales is quite compelling, this scenario would be subject to various constraints. 
On the other hand, if the scale is smaller than galactic size, then cosmological constraints would not directly apply, and more possibilities would be available.

In this work, we consider the dark matter to be made of very light particles. 
We consider masses $m_\phi\ll 1$\,eV; in this regime it is essential that the particle is a boson, so that one can be at high occupancy number. We will focus on scalar bosons. This can provide a type of cold dark matter due to formation from some kind of misalignment mechanism.
The most prominent example being the QCD axion \cite{Weinberg1978,Wilczek1978,Preskill1983,Abbott1983,Dine1983}. In some of the simplest QCD axion scenarios, the axion mass is $m_\phi\sim 10^{-5}-10^{-6}$\,eV.
Furthermore, interesting galactic behavior can take place if the axions are even much lighter than this; so-called ultralight. 
For such light axions, the number density must be correspondingly large to account for all of the dark matter.
In this case, the de Broglie wavelength $\lambda=h/(m\,v)$ can have astronomical consequences, such as leading to wave like behavior and altering the centers of galaxies \cite{Hu:2000ke}. The mass range around $m_\phi\sim 10^{-22}-10^{-19}$\,eV, or so, is often considered.
These ideas have been subject to various observational constraints, such as in Refs.~\cite{Schive:2015kza,dalal2022fuzzy, Ir_i__2017, Armengaud_2017, Chan_2022, Powell_2023,Hertzberg:2022vhk}. Furthermore, there has been significant interest in the case of multiple ultralight axions, such as in Refs.~\cite{Svrcek2006,Huang_2023, luu2023nested, Gosenca_2023, Glennon_2023, jain2023kinetic, Eby_2020, Amin_2022, vanDissel:2023vhu,Guo_2021}.
Related works has also explored solitons and halo phenomenology in scalar and vector dark matter with nonminimal couplings to gravity \cite{Chen:2024pyr, Zhang:2023fhs}.

One of the interesting properties of light axions is that under some circumstances they can organize into a kind of condensate; a ground state at fixed particle number. These are sometimes called ``boson stars" or ``clumps" and are a kind of ``soliton"; some explorations include Refs.~\cite{Chavanis:2011zi,Chavanis:2011zm,Liebling:2012fv,Schiappacasse:2017ham,Visinelli:2017ooc,Guth:2014hsa,Hertzberg:2018zte,Hertzberg:2020dbk, Zhang:2024bjo}. 
The formation and structure of these solutions are likely well described by classical field theory (discussions on aspects of this topic, include Refs.~\cite{Sikivie:2009qn,Hertzberg:2016tal,Dvali:2017ruz,Allali:2020ttz,Allali:2020shm,Allali:2021puy,Eberhardt:2022rcp,Cheong:2024ose}).
Simulations have indicated that these may form at the centers of galaxies
\cite{Schwabe:2016rze,Schive_20142,Bar_2018,Li_2021} or from mini-clusters and halos \cite{Eggemeier:2019jsu,Chen:2020cef} and reinforced by semi-analytical work \cite{Levkov_2018,dmitriev2024selfsimilar}. This is especially interesting as some observations hint that the galactic core density $\rho_c$ is related to the core radius $R_c$ by the relation $\rho_c\propto 1/R_c^q$, with $q\approx 1.3$; for some data, see Ref.~\cite{Rodrigues_2017}. 
However, it was shown in Ref.~\cite{Deng:2018jjz} that the ultralight axion proposal to explain cores of galaxies \cite{Hu:2000ke,Harko:2011xw,Robles:2012uy} does not fit the galactic data: If the ultralight axion interacts only (or primarily) via gravity then the relation predicted is very different, namely $q=4$. Furthermore, as shown in Ref.~\cite{Deng:2018jjz}, even if the scalar theory is altered by the inclusion of a local scalar potential $V(\phi)$, the problem persists.

In this work we examine light scalar (axion) dark matter that has a non-gravitational force. We take the mediator to also be a scalar with  mass $m_\chi$. We compute the properties of these condensate solutions in the presence of the new force. We find that the core density versus core radius relation is altered in way to make the relation less steep, as there is a transition in behavior from the $R_c\ll m_\chi^{-1}$ regime to the $R_c\gg m_\chi^{-1}$
regime. However, for moderate values of the coupling, we do not find that it can be increased all the way from $q=4$ to $q\approx1.3$, but rather a much more modest improvement. We also provide a first step towards the extension to multiple mediators. A full exploration of  parameter space, including large couplings, and detailed comparison to data is left for future work.

More generally, we view this work as a first step towards exploring new interesting phenomenology in the dark sector. Possible new forces between dark matter particles can lead to new phenomenology and astrophysical consequences. The effects on condensates are just one possibility, but are the focus of this initial study.

Our paper is organized as follows: In Section \ref{theory} we introduce the model and construct the equations of motion in the appropriate nonrelativistic limit. 
In Section \ref{results} we numerically solve the equations to obtain the soliton solutions.
In Section \ref{Multi} we extend the analysis to the multiple mediator case; using two mediators as an example.
Finally, in Section \ref{discussion} we discuss our findings and mention possible future work.

\section{Theory of Scalar Dark Matter with New Force}\label{theory}

Let us consider the dark matter as a scalar boson $\phi$ with mass $m_\phi$. We assume it is minimally coupled to gravity. In addition, we introduce a new light mediator $\chi$ with mass $m_\chi$. To leading order, we assume that $\chi$ couples to $\phi$ via  3-point interactions of the form $\chi\,\phi^2$ and $\chi\,(\partial\phi)^2$. In this way, it can mediate a new force between $\phi$ particles. 

The action is taken to be of the form
\beq
    S= \int d^4 x \sqrt{-g}\left[ \Lagr_\phi + \Lagr_\chi +  \Lagr_{\rm int} + \Lagr_{\rm grav}\right]\,,
    \label{eq1}
\eeq
where each Lagrangian density can be explicitly written as (we use units $\hbar=c=1$ and metric signature $+ - - - $)
\begin{align}
       \Lagr_\phi &= \frac{1}{2} g^{\mu \nu} \partial_\mu \phi \partial_\nu \phi -V(\phi)\,,  \label{eq3} \\[5 pt]
       \Lagr_\chi &= \frac{1}{2} g^{\mu \nu} \partial_\mu \chi \partial_\nu \chi - \frac{1}{2} m_\chi^2 \chi^2\,, \label{eq4} \\[5 pt]
        \Lagr_{\rm int} &= -\chi \left[ c_1 g^{\mu \nu} \partial_\mu \phi \partial_\nu \phi + c_2 \tilde{V}(\phi) \right] \label{eq6b}\,, \\[5 pt]
        \Lagr_{\rm grav} &=   -\frac{\mathcal{R}}{16 \pi G} \,. \label{eq5} 
\end{align}
Here $\mathcal{R}$ is the Ricci scalar, $G$ is Newton's gravitational constant. 
Importantly, we have introduced a pair of coupling constants: $c_1$, which is the strength of $\chi$ coupling to the kinetic term, and $c_2$, which is the strength of coupling to the potential term.
One could go further and have higher order interactions, like $c_1^2\chi^2(\partial\phi)^2,\,c_2^2\chi^2\tilde{V}(\phi)$, etc. In fact a tower of operators controlled by $c_{1,2}\,\chi$ is possible. 
To suppress such higher order terms, we work in the regime $c_{1,2}\,\chi\ll1$.

The potential of the scalar is assumed to have an expansion around $\phi=0$ of the form
\beq
    V(\phi) = \frac{1}{2} m_\phi^2 \,\phi^2 + \ldots
    \label{eq6}
\eeq
Higher order terms $\sim\phi^3$ or $\phi^4$ can be included in principle. However, in this work, we shall focus on the small field regime, where the quadratic mass term $m_\phi^2\phi^2/2$ dominates. This is often the case for dark matter in the galaxy today, since the field has undergone significant red-shifting from the early universe. Although cases in which the nonlinear corrections matter are conceivable and worth considering too. In the case of an axion, one considers periodic potentials of the form
\beq
V(\phi)=m_\phi^2 f_{\rm PQ}^2(1-\cos(\phi/f_{\rm PQ}))\,,
\eeq
where $f_{\rm PQ}$ sets the scale of a spontaneously broken $U(1)$ symmetry at high energies.
In the case of the QCD axion \cite{Peccei1977}, nontrivial dynamics associated with QCD instantons generate $\mathcal{O}(1)$ corrections to this shape, but the details are not essential here.
While it may be natural to have $\phi\sim f_{\rm PQ}$ in the early universe, smaller values $\phi\ll f_{\rm PQ}$ can be natural in the late universe, where the quadratic approximation can be accurate (although exceptions are conceivable). 

Finally, there is the choice of the potential term $\tilde{V}(\phi)$ that appears in the interaction $\sim\chi\,\tilde{V}(\phi)$. A common consideration is to imagine that the scalar $\chi$ couples to the trace  of the energy-momentum tensor of the $\phi$ field, which is
\beq
T_{\mu\nu}^\phi=
\partial_\mu\phi\partial_\nu\phi
-g_{\mu\nu}\left(
\frac{1}{2} g^{\alpha \beta} \partial_\alpha \phi \partial_\beta \phi -V(\phi)\right)\,.
\eeq 
Its trace is
\beq
T^\phi=-2\left({1\over2}g^{\mu\nu}\partial_\mu\phi\partial_\nu\phi
-2 V(\phi)\right)\,.
\eeq
This form of the coupling $\sim\chi\,T^\phi$ can be achieved by taking
\beq
\tilde{V}(\phi)=V(\phi)\,,
\eeq
and taking $c_2=-4c_1$. In this case, as $\phi$ oscillates, then the source $T^\phi$ will oscillate (pressure oscillations). This is a very interesting possibility, but will not be the focus of the present work. Another possibility is that $\chi$ is primarily sourced by the energy density $\rho_\phi$, which is only slowly varying for cold dark matter. We see that again if we take $\tilde{V}=V$, and we pick $c_1=c_2/2$, then $\chi$ is approximately sourced by the energy density. We shall return to this scenario shortly.

\subsection{Weak Field Regime}

In this work, we will focus on the weak field regime and will study situations of negligible gravitational waves. This is appropriate for dynamics in the galaxy. The corresponding metric can be taken to be of the form
\beq
    ds^2=(1+2 \phi_N) dt^2 - (1-2 \psi) d\Vec{x}\cdot d\Vec{x} \,,
    \label{eq1.5}
\eeq
where $\phi_N, \,\psi$ are of the same order and not necessarily equal, unless we impose the static limit. 


We expand the Einstein-Hilbert term to second order, to obtain the gravitational term as (ignoring total derivatives)
\beq
    -\sqrt{-g}\,\mathcal{R} = 2[(\Vec{\nabla} \psi)^2 -3 \dot{\psi}^2 - 2\Vec{\nabla} \phi_N \cdot \Vec{\nabla} \psi ] \,.
    \label{eq8}
\eeq
Then the full Lagrangian density becomes
\beq
    \begin{aligned}
        \Lagr_{\rm full} = &\frac{1}{2} (\partial_\mu\phi)^2 - V(\phi) + \frac{1}{2} (\partial_\mu\chi)^2 - \frac{1}{2} m_\chi^2\chi^2 
        -\phi_N \left[ \frac{1}{2}\dot{\phi}^2+\frac{1}{2}( \vec{ \nabla} \phi)^2  +V(\phi) \right]  \\
        -&\psi \left[ \frac{3}{2}\dot{\phi}^2-\frac{1}{2}( \vec{ \nabla} \phi)^2  - 3V(\phi) \right]  
        - \chi\left[ c_1 (\dot{\phi}^2 - ( \vec{ \nabla} \phi)^2)  + c_2\tilde{V}(\phi) \right]  \\
        +& \frac{1}{8 \pi G}[(\vec{\nabla} \psi)^2 -3 \dot{\psi}^2 - 2\vec{\nabla} \phi_N \cdot \vec{\nabla} \psi ]\,,
    \end{aligned}
    \label{eq9}
\eeq
where we have expanded to linear order in $\phi_N,\,\psi,\,\chi$ in the interaction terms. Since we focus here on the weak field regime (meaning $\phi_N\ll1,\,\psi\ll1$ (units $c=1$), along with $c_{1,2}\,\chi\ll1$), higher order corrections are suppressed.


\subsection{Nonrelativistic Regime}

We are interested in the nonrelativistic regime for cold dark matter in the galaxy.
To analyze this, it is useful to perform a decomposition of the dark matter field $\phi$ as follows
\beq
    \phi(t,\vec{x})= \frac{1}{\sqrt{2m_\phi}} [e^{-i m_\phi t} F(t,\vec{x}) + e^{i m_\phi t} F^*(t,\vec{x})]\,. \label{eq10}
\eeq
Here we have factorized for the rapid variation at the particle mass frequency $m_\phi$, and captured the remaining slow variation in the function $F$.
We now insert this into $\mathcal{L}_{\rm full}$ and time average over any rapid oscillations. This is valid when the characteristic wavelengths of the dark matter is much longer than the particle's Compton wavelength $m_\phi^{-1}$ (which is indeed the nonrelativistic regime). Corrections to this are suppressed by factors of $v^2\ll1$ (in units $c=1$), as they are relativistic corrections.
For work on including relativistic corrections systematically, see Ref.~\cite{Namjoo:2017nia, Salehian:2021khb}.
We also take $\tilde{V}=V$ and work in the small field regime with $V(\phi)\approx m_\phi^2\phi^2/2$.

Varying the action with respect to $\phi_N$, one generates the standard result that $\psi$ obeys the Poisson equation
\beq
\nabla^2\psi=4\pi G\,\rho_\phi\,,
\eeq
where in this limit, the energy (mass) density associated with $\phi$ is
\beq
\rho_\phi=m_\phi |F|^2\,,
\eeq
with $|F|^2$ the number density of $\phi$-particles. So self-consistently, $\psi$ must be slowly varying and we can ignore the $\dot\psi$ terms in the above Lagrangian. 
We can then time average the term multiplying $\psi$ in the Lagrangian which then vanishes to leading order. Then varying the action with respect to $\psi$, we obtain 
\beq
\phi_N=\psi\,,
\eeq
as is standard.

The corresponding Lagrangian, after dispensing with the pieces that time average to zero, is
\beq
    \begin{aligned}
        \Lagr_{\rm time \: avg} = &\frac{i}{2} (\dot{F}F^*-F\dot{F}^*) - \frac{1}{2m_\phi}|\vec{\nabla}F|^2  
        -\phi_N \, \rho_\phi -\frac{1}{8 \pi G}(\vec{ \nabla} \phi_N)^2 \\
        &+ \frac{1}{2}\dot{\chi}^2-\frac{1}{2}(\vec{ \nabla} \chi)^2  - \frac{1}{2}m_\chi^2\chi^2 - C\,\chi\, \rho_\phi +
        {m_\phi\over 2}\bar{C}\left(e^{-2im_\phi t}\,\chi\,F^2+c.c.\right)\,,
    \end{aligned}
    \label{eq12}
\eeq
where
\beq
C\equiv c_1+{c_2\over 2},\,\,\,\,\,
\bar{C}\equiv c_1-{c_2\over 2}.
\eeq

In the final term in Eq.~(\ref{eq12}), we have not time averaged the fast oscillations $e^{-2im_\phi t}$ to zero, as it is possible that $\chi$ also oscillates rapidly and the product may vary slowly. This can arise from the production of $\chi$ waves from an oscillating $\phi$ solution. We can consider this in future work, but it is not our focus here. Instead we consider the simpler case in which $\chi$ is also slowly varying and only acts as a force mediator. This will occur if 
$\bar{C}$ vanishes (or is small). So we consider the special case
\beq
c_1={c_2\over 2}\,,
\eeq
which sets $\bar{C}=0$. For $c_1\neq c_2/2$ there is a possibility of $\chi$ waves, which may be a small effect because the couplings will be taken to be only of order gravitational strength. But it is left for future work for its details. In any case, we ignore this possibility in this work, and take $\chi$ to be slowly varying too. This means the final term is ignored and so too is the $\dot\chi^2$ term. In this case, the Lagrangian reduces to
\beq
    \begin{aligned}
        \Lagr_{\rm time \: avg} = &\frac{i}{2} (\dot{F}F^*-F\dot{F}^*) - \frac{1}{2m_\phi}|\vec{\nabla}F|^2  
        -\phi_N \, \rho_\phi -\frac{1}{8 \pi G}(\vec{ \nabla} \phi_N)^2 \\
        &-\frac{1}{2}(\vec{ \nabla} \chi)^2  -{1\over2}m_\chi^2\chi^2-C\,\chi\,\rho_\phi\,.
    \end{aligned}
    \label{Lsimple}
\eeq

\subsection{Equations of Motion}

Let us now vary the action with respect to the dark matter field $F$. This gives the 
nonrelativistic equation of motion
\beq
    i \,\dot{F} = - \frac{\nabla^2 F}{2 m_\phi} + m_\phi \left( \phi_N + C\, \chi \right) F\,.
    \label{eq12.5}
\eeq
The constraint equation for $\phi_N$ is the usual Poisson equation:
\beq
    \nabla^2 \phi_N = 4 \pi G \rho_\phi\,.
    \label{eq13}
\eeq
And the constraint equation (having ignored its dynamics in this limit)  for the new force mediator $\chi$ is a screened Poisson equation
\beq
     \nabla^2 \chi - m_\chi^2 \chi = C\, \rho_\phi\,.
    \label{eq15}
\eeq

\subsection{Static and Spherically Symmetric Configurations}

The above set of 3 equations can in principle be solved numerically to study the behavior of dark matter in the galaxy (ignoring baryons). However, in this work we focus on condensate solutions; configurations that minimize the energy at fixed particle number.

These solutions have pure phase time dependence as
\beq
F(t,\vec{x})=e^{-i\,\mu\, t}\,f(\vec{x})\,,
\eeq
where $\mu$ is a kind of chemical potential and $f(\vec{x})$ is a function of space only; which we can take to be real without loss of generality. The corresponding density $\rho_\phi=m_\phi|F|^2=m_\phi\,f^2$ is therefore static and is a kind of soliton.

In the case of a pure gravitational interaction, it is known that the ground state configurations are spherically symmetric. We anticipate this remains true in the presence of this new $\chi$ force.
Assuming spherical symmetry then, we can replace the Laplacian operator with
\beq
\nabla^2\to{\partial^2\over \partial r^2}+{2\over r}{\partial\over \partial r}\,.
\eeq
Then $f=f(r)$ obeys the following ordinary differential equation
\beq
    \mu\,f = - {1\over 2m_\phi}\left(f''+{2\over r}f'\right) + m_\phi \left( \phi_N + C\, \chi \right) f\,,
    \label{fstaticsp}
\eeq
where primes mean derivatives with respect to radius $r$.
Using the spherical symmetry, we can integrate up the Poisson equation to obtain the Newton potential as a 1-dimensional radial integral
\beq
    \phi_N(r) = -{4\pi G m_\phi\over r} \int_0^\infty dr' \,r'\,r_<\,f(r')^2 +\mbox{const}\,,
\eeq
where  $r_< \equiv \min \{r, r'\}$ and we have allowed for the possibility of a constant shift depending on boundary conditions. 

For the mediator $\chi$, we note that in the massless case $m_\chi=0$ then $\chi$ also obeys the Poisson equation. This means it is proportional to $\phi_N$ (possibly up to a constant), as
\beq
\chi=\frac{C}{4 \pi G}\,\phi_N+\mbox{const}\,\,\,\,\,\,(m_\chi=0)\,.
\eeq 
In the massive case $m_\chi\neq0$ (which is our focus in this work), the solution is rather more involved. However, it has an integral representation in terms of a Yukawa type Green's function. Due to the spherical symmetry, it can be reduced to the following 1-dimensional radial integral
\beq
    \chi(r) = -{C m_\phi\over r} \int_0^\infty dr' \,\frac{r' \,f(r')^2}{2 m_{\chi}} e^{- m_\chi
    (r+r')} [e^{2 m_\chi r_<}-1]\,. \label{eq16} 
\eeq
Since $\chi$ is massive, we must demand $\chi\to 0$ as $r\to\infty$, so there is no constant shift.

\subsection{Defining Dimensionless Quantities}

The above set of variables can be reduced to dimensionless
quantities. To do so, we define
    \begin{align}
        f(r) &\equiv  \alpha \:  \tilde{f}(r) \,, \\[5 pt]
        \phi_N(r) &\equiv \gamma \: \tilde{\phi}_N (r) \, ,\\[5 pt]
           \chi(r)&\equiv{\gamma\over C}\,\tilde{\chi}(r)\,,\\[5 pt]
                r &\equiv \beta \: \tilde{r} \, , \\[5 pt]
        \mu &\equiv \varepsilon \: \tilde{\mu} \, ,
            \label{eq17}
    \end{align} 
    where $\alpha,\,\gamma,\,\beta,\,\varepsilon$ are constant conversion factors and $\tilde{f},\,\tilde{\phi}_N,\,\tilde{\chi},\,\tilde{r},\,\tilde{\mu}$ are dimensionless variables. 
Using the equations for the fields in Eqs.~\ref{fstaticsp}, \ref{eq13}, \ref{eq15} and these dimensionless quantities, we choose the following expressions for the scaling factors in terms of $\alpha$ parameter:
\begin{align}
        \beta &=  (G\, m_\phi^3\, \alpha^2)^{-1/4}  \,, \\[5 pt]
        \gamma &= \sqrt{G\,\alpha^2\over m_\phi}\, , \\[5 pt]
        \varepsilon &=  m_\phi\gamma \, .
 \label{eq18}   \end{align} 
We choose the parameter $\alpha$ to correspond to the value of the field  at the center, i.e.,
\beq
\alpha = f(0)\,.
\eeq
So by definition, we have $\tilde{f}(0)=1$.
Then the central (or ``core") density $\rho_c=\rho(0)$ can be expressed as
\beq
\rho_c = m_\phi\, \alpha^2 = \frac{1}{G\,m_\phi^2\,\beta^4} \,.
    \label{eq19}
\eeq
In the upcoming plots, we shall display the dimensionless field value $\tilde{f}$ versus dimensionless radius $\tilde{r}$. Using the above, these are just a re-writing of the physical density and radius as
\beq
\tilde{f}=\sqrt{\rho\over\rho_c},
\,\,\,\,\,\tilde{r} = r\,(G\,m_\phi^2\,\rho_c)^{1/4}\,.
\eeq
 
Now we can express the field equations using these dimensionless parameters 
    \begin{align}
        \tilde{\mu}\, \tilde{f} &= - \frac{\tilde{\nabla}^2}{2} \tilde{f} + (\tilde{\phi}_N + \tilde{\chi})\tilde{f} \, , \label{feqdim}\\
             \tilde{\nabla}^2 \tilde{\phi}_N &= 4\pi \tilde{f}^2\, ,\label{phieqdim}\\
        \tilde{\nabla}^2 \tilde{\chi} - (\beta')^2 \tilde{\chi} &=  4 \pi g_{\chi} \tilde{f}^2 \, .
    \label{chieqdim}    \end{align}
Here we have identified that there are {\em two} physical parameters in the system that cannot be scaled out. They are a dimensionless coupling constant $g_\chi$ and a dimensionless mass of the mediator $\beta'$, defined by 
    \begin{align}
        g_\chi &\equiv \frac{ C^2}{4 \pi G} \, , \\[5 pt]
        \beta' &\equiv m_\chi \beta \, .
           \label{eq21}
    \end{align}
 Here $g_\chi$ measures how strong the new force is relative to the gravitational force.
 And $\beta'$ is a measure of the characteristic radial size of the solution relative to the Compton wavelength of the mediator.

\subsection{Effective Gravitational Coupling}

We note that there are 2 important limits to consider: (i) For $\beta'\ll 1$, then the $\chi$ field is effectively massless and it acts as a new long range mediator. In this case, we have $\chi\approx C\phi_N/(4\pi G)+$const, and the effective strength of gravitation is enhanced by this new $\chi$ mediated force to
\beq
    G_{\text{eff}}=(1+g_\chi)\,G,\,\,\,\,\,\beta'\ll1\,.
    \label{eq22}
\eeq
(ii) For $\beta'\gg 1$, then the $\chi$ field is effectively very heavy and its effects are exponentially suppressed. In this case the effective gravitational constant is just 
\beq
G_{\rm eff}=G,\,\,\,\,\,\beta'\gg1\,,
\eeq
as usual. In the intermediate case $\beta'\sim 1$, then we have a non-trivial new force, which must be handled numerically.

Using the above relationships, we see that the central density is related to the characteristic radius, or core radius $R_c$, as
\beq
\rho_c = {\zeta\over G_{\rm eff}\,m_\phi^2\,R_c^4}\,,
\label{rhoRrel}\eeq
where $\zeta$ is an $\mathcal{O}(1)$ number, which comes from carrying out the numerics and depends on the precise definition of the core radius.
Here we have noted that $\beta$ sets the characteristic size, so $\beta\sim R_c$. Furthermore, we have corrected by the appropriate $G_{\rm eff}$.


\section{Numerical Results}\label{results}

\subsection{Numerical Procedure}

We have solved the above system of equations to obtain the condensate solutions. To do so, we employ a perturbative procedure as follows: 
\begin{enumerate}[{(i)}]
\item We first set $\tilde{\chi}=0$.
    \item We solve the $\{\tilde{f},\,\tilde{\phi}_N\}$ system of equations (\ref{feqdim},\,\ref{phieqdim}), subject to the boundary conditions 
\beq
\tilde{f}(0)=1,\,\,\,\,
\tilde{f}'(0)=0,\,\,\,\, 
\tilde{\phi}_N(0)=0,\,\,\,\,
\tilde{\phi}'_N(0)=0.
\eeq
To do so, we first make an initial guess for the chemical potential $\tilde{\mu}$. But a random value will not give the soliton solution. Instead it will typically give a diverging solution at large $\tilde{r}$. So we keep altering the value of $\tilde{\mu}$ either higher or lower, so as to make the solution for $\tilde{f}$ monotonically decrease toward zero at large $\tilde{r}$.
We rapidly achieve convergence of $\tilde{\mu}$ by using the method of bisecting the interval of $\tilde{\mu}$. 
\item We then use this solution for $\tilde{f}$ and feed it into the integral representation for $\tilde{\chi}$; the dimensionless version of Eq.~(\ref{eq16}), i.e.,
\beq
    \tilde{\chi}(\tilde{r}) = -{4\pi g_\chi \over \tilde{r}} \int_0^\infty d\tilde{r}' \,\frac{\tilde{r}' \,\tilde{f}(\tilde{r}')^2}{2 \beta'} e^{- \beta'
    (\tilde{r}+\tilde{r}')} [e^{2 \beta' \tilde{r}_<}-1]\,. \label{chidimintegral} 
\eeq
\item Using this new value for $\tilde{\chi}$, we repeat the above procedure in (ii) and (iii) several times until we achieve convergence for $\{\tilde{f},\,\tilde{\phi}_N,\,\tilde{\chi}\}$.
\end{enumerate}
We note that a limitation of this procedure is that it is inefficient if $g_{\chi}$ is large. As stated above, the procedure begins with setting $\tilde{\chi}=0$ and then solving recursively. This converges rapidly if $g_{\chi}$ is small as the perturbed $\tilde{\chi}$ is small. In fact we find this procedure is still somewhat rapid even if $g_{\chi}\sim 1$ (and in fact we take $g_{\chi}=1$ and $g_{\chi}=2$ in several of our upcoming plots). But its convergence is quite slow for $g_\chi \gg 1$; we do not focus on that scenario here, but leave for future work.

\subsection{Results}
\begin{figure}[h!]
    \centering
        \includegraphics[width=0.49\textwidth,height=6.2cm]{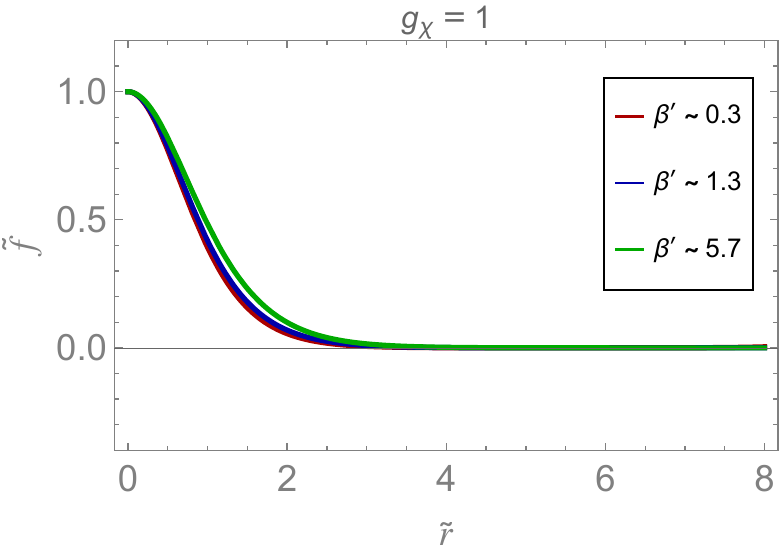}\,
        \includegraphics[width=0.49\textwidth,height=6.2cm]{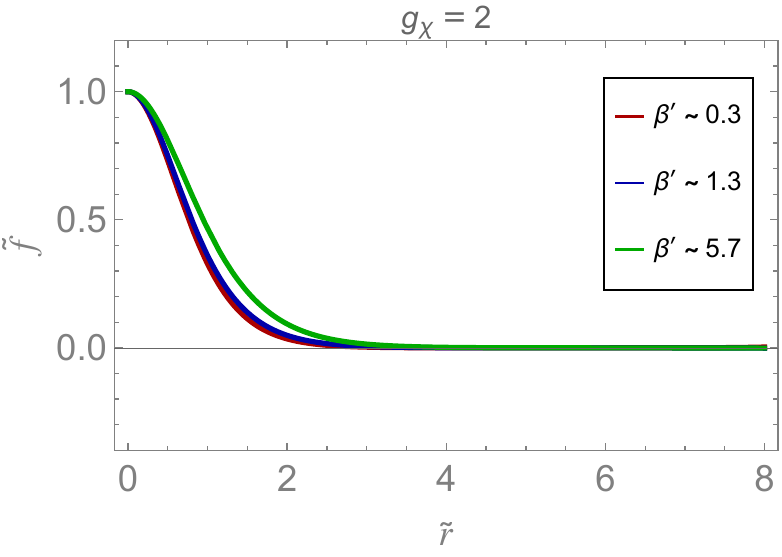}\\
        \vspace{0.2cm}
       \includegraphics[width=0.49\textwidth,height=6.2cm]{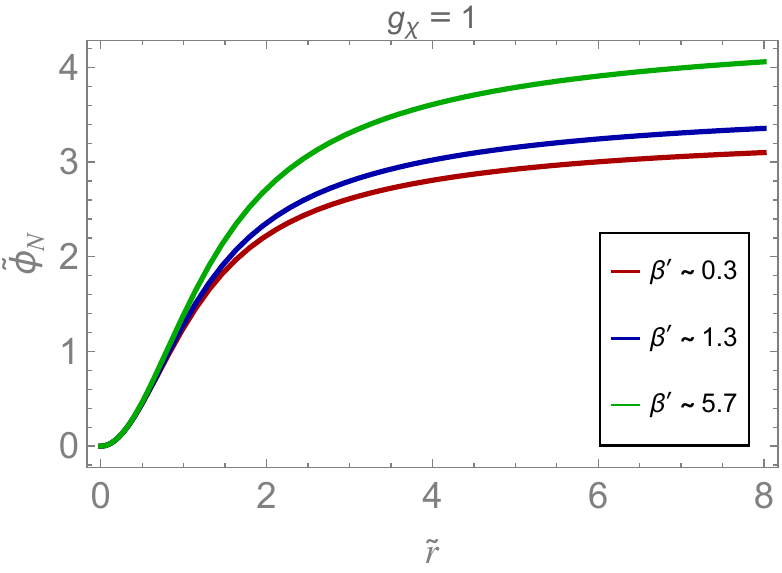}\,
        \includegraphics[width=0.49\textwidth,height=6.2cm]{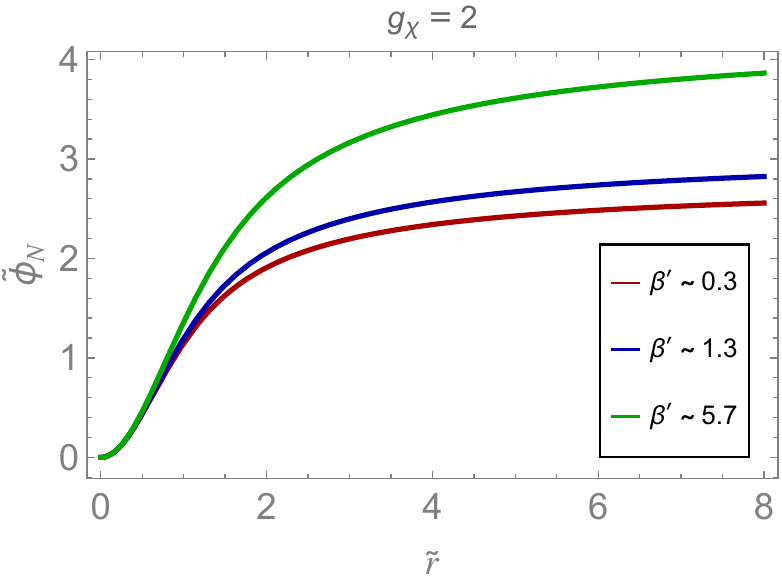}\\
        \vspace{0.2cm}
        \includegraphics[width=0.49\textwidth,height=6.2cm]{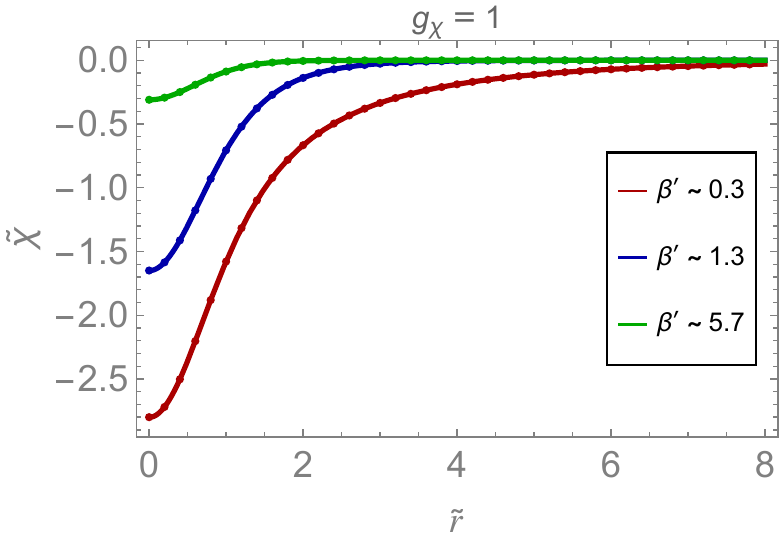}\,
        \includegraphics[width=0.49\textwidth,height=6.2cm]{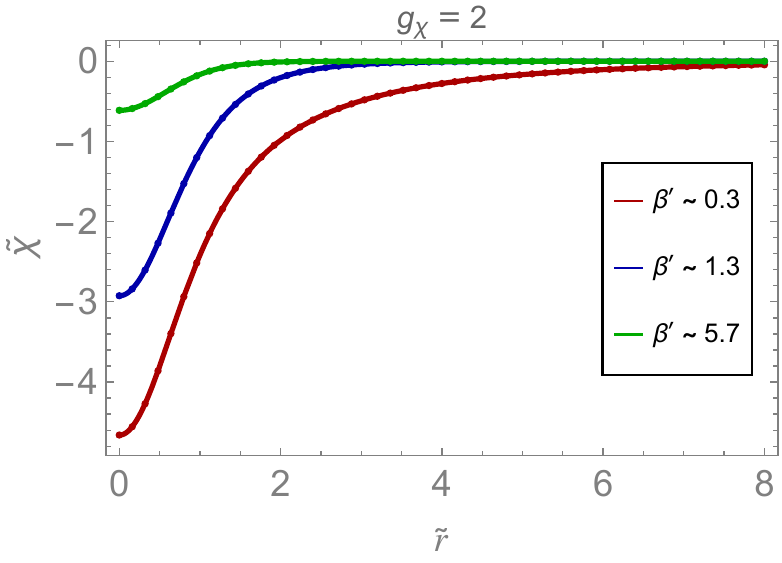}
    \caption{ Representative plots of fields $\tilde{f}, \tilde{\phi}_N , \tilde{\chi}$ versus dimensionless radius $\tilde{r}$ evaluated at $\beta' = 0.287 \sim 0.3$, $\beta' = 1.278 \sim 1.3$ and $\beta' = 5.683 \sim 5.7$.
    In the left (right) column, we have set the coupling of the new force to be $g_\chi=1$ ($g_\chi=2)$. }
    \label{fig:repplot-f}
\end{figure}

 In Figure \ref{fig:repplot-f} we show some representative plots of the field $f$ versus radius, as well as the Newton potential and the $\chi$ potential versus radius. Naturally we see that for large $\beta'$ (i.e., large $m_\chi$) the $\chi$ field is exponentially suppressed and unimportant. While for small $\beta'$ (i.e., small $m_\chi$) the $\chi$ field is appreciable and important.

From the numerical solution for the soliton, we can define a type of core radius $R_c$. We define this by the full width half maximum, i.e., the value of $r=R_c$ such that
\beq
f(R_c)^2={1\over2}f(0)^2\,.
\eeq
Or, in terms of the dimensionless $\tilde{f}$, this is $\tilde{f}(R_c)^2=1/2$ (as $\tilde{f}(0)=1$). 

\begin{figure}[h!]
    \centering
    \includegraphics[width=0.49\linewidth,height=6.2cm]{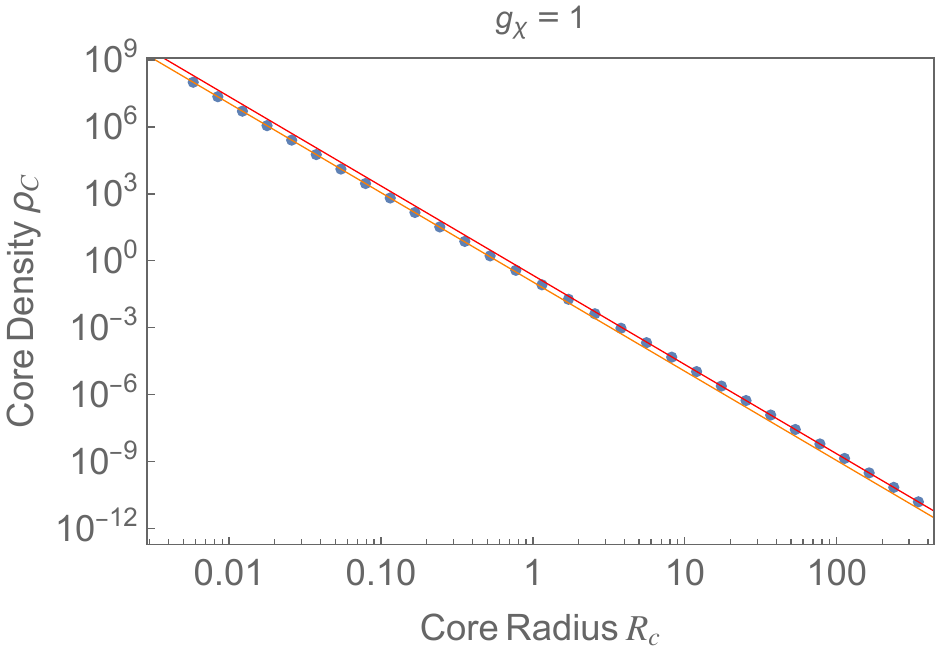}\,
      \includegraphics[width=0.49\linewidth,height=6.2cm]{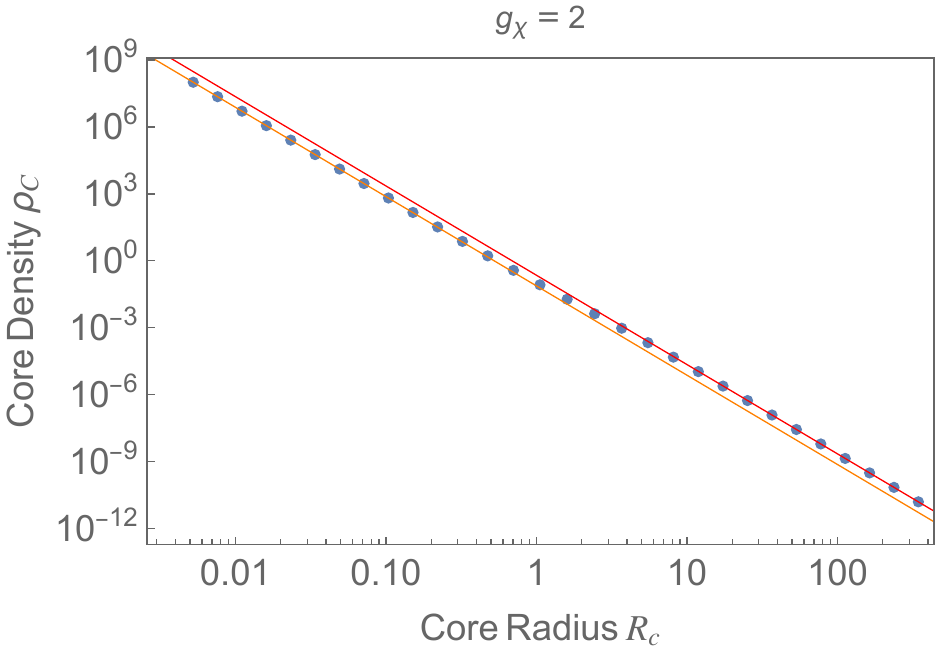}\\
      \vspace{0.2cm}
      \includegraphics[width=0.49\linewidth,height=6.2cm]{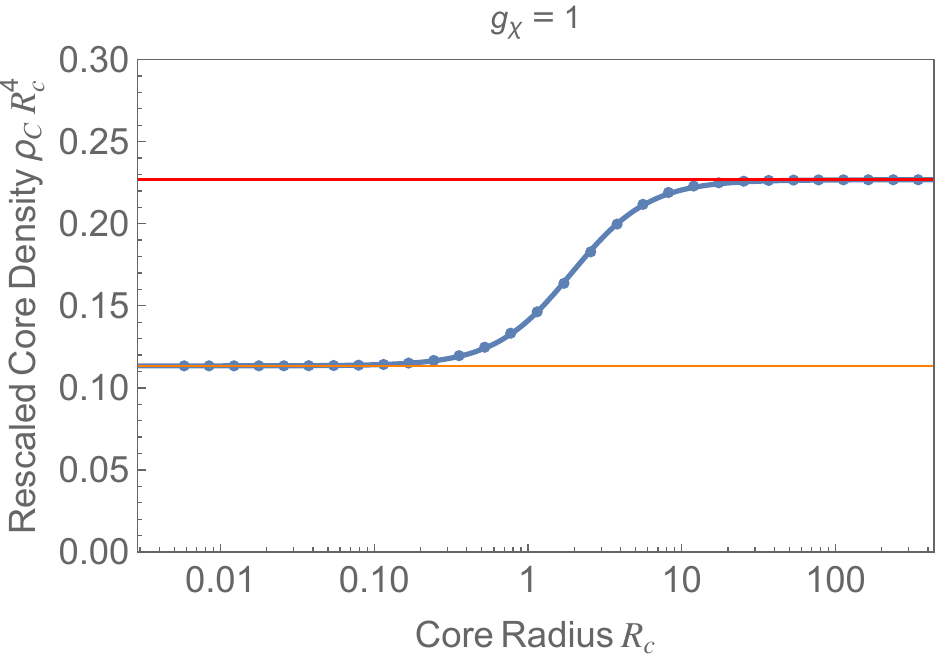}\,
      \includegraphics[width=0.49\linewidth,height=6.2cm]{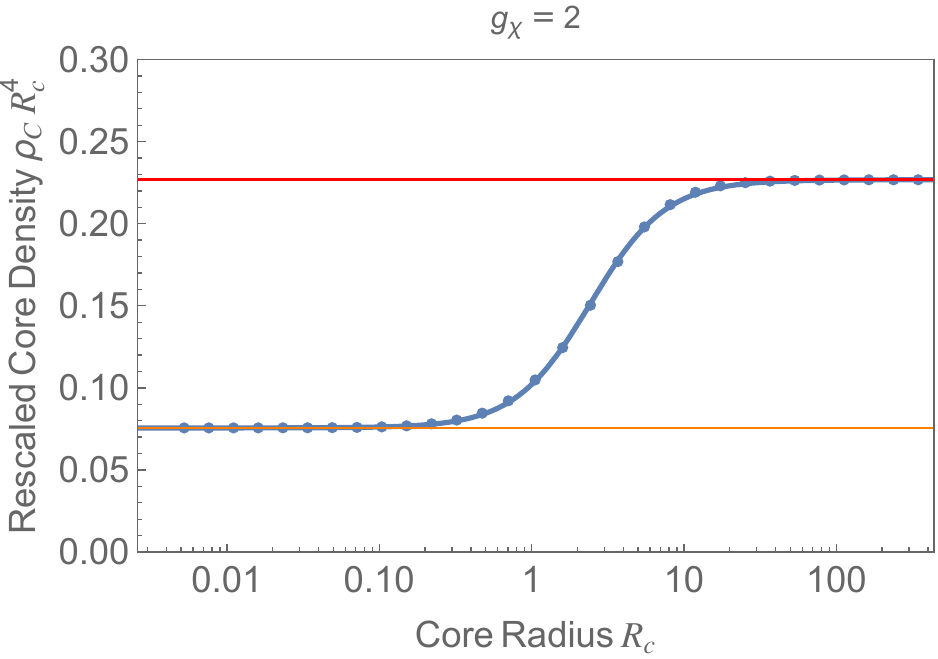}
    \caption{Top: Core density $\rho_c$ (in units of $m_\chi^4/(G \,m_\phi^2)$) of the soliton versus core radius $R_c$ (in units of $m_\chi^{-1}$).  
    Bottom: Rescaled core density $\rho_c\, R_c^4$ (in units of $(G\, m_\phi^2)^{-1}$) of the soliton versus core radius $R_c$ (in units of $m_\chi^{-1}$). 
    Small $R_c\ll m_\chi^{-1}$ and large $R_c\gg m_\chi^{-1}$ asymptotes are given in orange and red lines, respectively. 
    Also a fit function of the form in Eq.~(\ref{fit}) is given in the bottom plot.
    In the left (right) column, we have set the coupling of the new force to be $g_\chi=1$ ($g_\chi=2)$ .} 
    \label{fig:rhoRrelation}
\end{figure}

In Figure \ref{fig:rhoRrelation} we plot the core density $\rho_c$ versus core radius $R_c$. For comparison, we also show the asymptotic predictions given in Eq.~(\ref{rhoRrel}). Using the above  definition of $R_c$, we numerically find that the $\mathcal{O}(1)$ prefactor  in Eq.~(\ref{rhoRrel}) is
\beq
\zeta\approx 0.226\,.
\eeq

In the bottom plot of Figure \ref{fig:rhoRrelation}, we plot the rescaled density $\rho_c\,R_c^4$, which shows the transition between these two regimes more clearly. For $R_c\ll m_\chi^{-1}$, both forces (gravity plus the new $\chi$ force) are important and the effective Newton's constant is enhanced to $G_{\rm eff}=(1+g_\chi)G$. While for $R_c\gg m_\chi^{-1}$, only gravity is important and the effective Newton's constant is  $G_{\rm eff}=G$.
In the left column, we have taken $g_\chi=1$ in this plot, so the former asymptote is $1/(1+1)=1/2$ times the latter asymptote.
While in the right column, we have taken $g_\chi=2$ in this plot, so the former asymptote is $1/(1+2)=1/3$ times the latter asymptote.

Furthermore, to compare to the numerical result, we have included an interpolating function between these 2 asymptotes. We have chosen the following form for the fit:
\beq
(\rho_c\,R_c^4)_{\rm fit} = {\zeta\over (1+g_\chi)G\,m_\phi^2}
\left(1+{g_\chi\over 1+a\,(R_c\,m_\chi)^{-p}}\right)\,,
\label{fit}\eeq
where $a>0$ and $p>0$ are dimensionless numbers determined from the best fit (which are themselves functions of $g_\chi$). We see from the figure that it provides a rather accurate fit to the numerical result.

\section{Multiple Force Mediators}\label{Multi}

We can also consider the more general case in which there are multiple force mediators. For the case of 2 mediators, $\chi_1$ and $\chi_2$, we can update our starting Lagrangian to
\begin{align}
       \Lagr_\chi &= \frac{1}{2} g^{\mu \nu} \partial_\mu \chi_1 \partial_\nu \chi_1 - \frac{1}{2} m_{\chi_1}^2 \chi_1^2
       +\frac{1}{2} g^{\mu \nu} \partial_\mu \chi_2 \partial_\nu \chi_2 - \frac{1}{2} m_{\chi_2}^2 \chi_2^2\,, \label{eq4} \\[5 pt]
        \Lagr_{\rm int} &= -\chi_1 \left[ c_1 g^{\mu \nu} \partial_\mu \phi \partial_\nu \phi + c_2 \tilde{V}(\phi) \right] 
        -\chi_2 \left[ \hat{c}_1 g^{\mu \nu} \partial_\mu \phi \partial_\nu \phi + \hat{c}_2 \tilde{V}(\phi) \right] \,.
        \label{twomediator} 
\end{align}
We then carry out the same procedure as before to take the weak field nonrelativistic limit, with $c_1=c_2/2$ and $\hat{c}_1=\hat{c}_2/2$ for each of the $\chi$ fields. This leads to the following equation of motion for the matter field
\beq
        i \,\dot{F} = - \frac{\nabla^2 F}{2 m_\phi} + m_\phi \left( \phi_N + C_1\, \chi_1+C_2\, \chi_2 \right) F\,,
            \label{F2}
        \eeq
and the following constraint equations for both force mediators
\begin{align}
         \nabla^2 \chi_1 - m_{\chi_1}^2 \chi_1 &= C_1\, \rho_\phi\,,\\
              \nabla^2 \chi_2 - m_{\chi_2}^2 \chi_2 &= C_2\, \rho_\phi\,,
\end{align}
with $C_1=c_1+c_2/2=c_2$ and $C_2=\hat{c}_1+\hat{c}_2/2=\hat{c}_2$.
Also $\phi_N$ still obeys the standard Poisson equation (\ref{eq13}).

\begin{figure}[t!]
    \centering
    \includegraphics[width=0.49\linewidth,height=6.2cm]{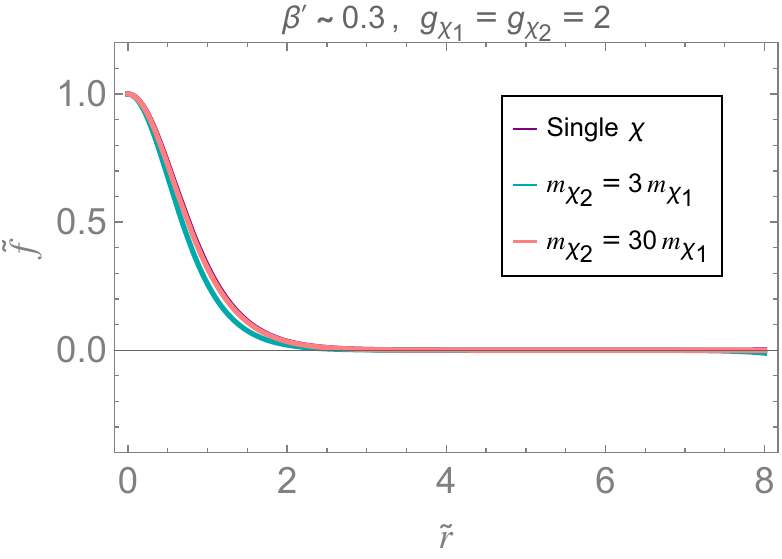}\,
      \includegraphics[width=0.49\linewidth,height=6.2cm]{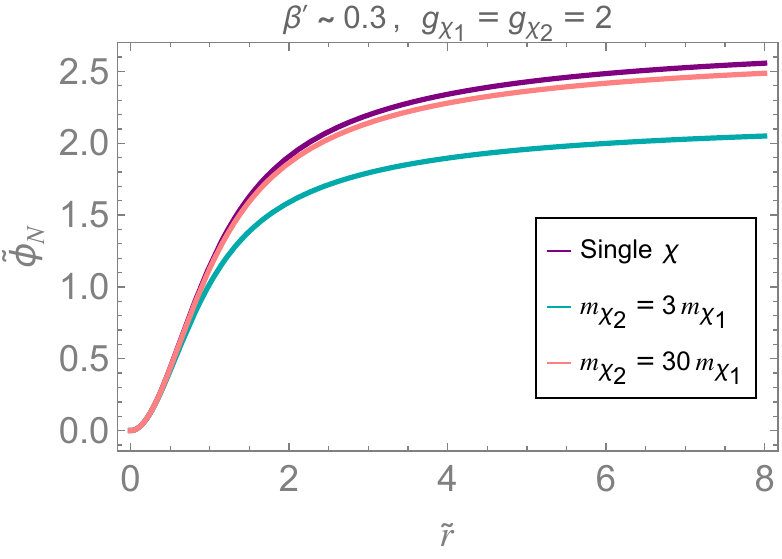}\\
      \vspace{0.2cm}
      \includegraphics[width=0.49\linewidth,height=6.2cm]{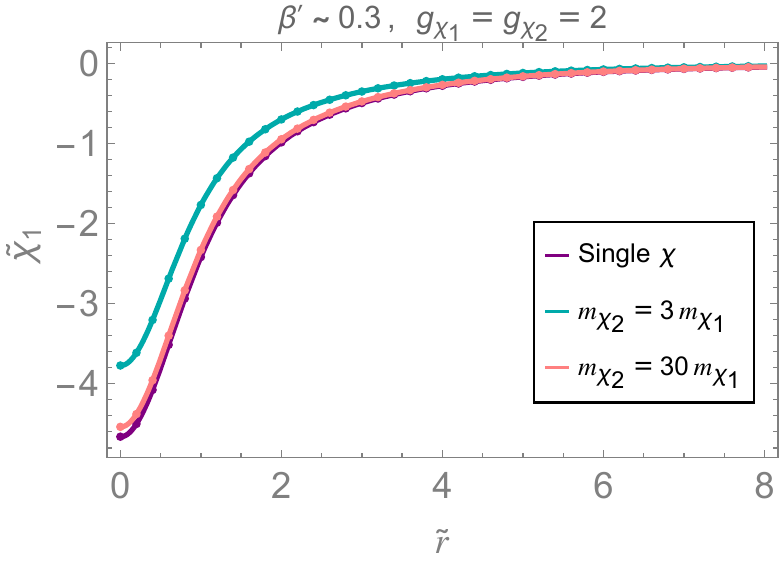}\,
      \includegraphics[width=0.49\linewidth,height=6.2cm]{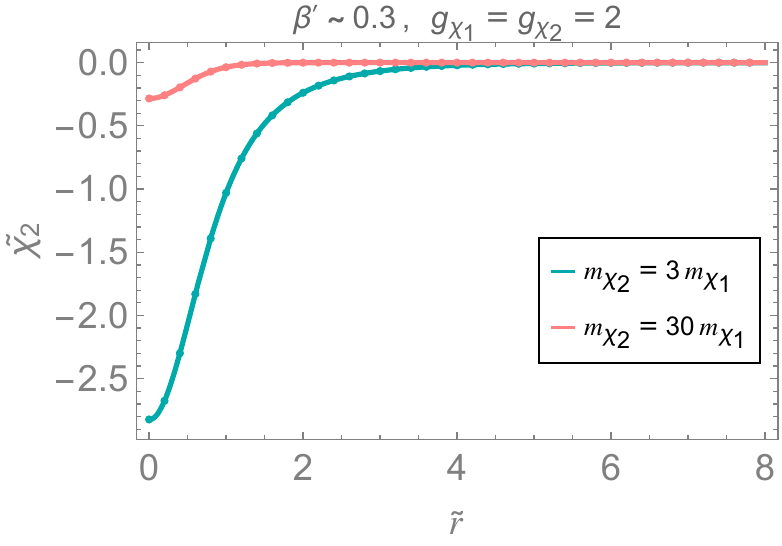}\\
      \vspace{-0.1cm}
\includegraphics[width=0.49\linewidth,height=6.2cm]{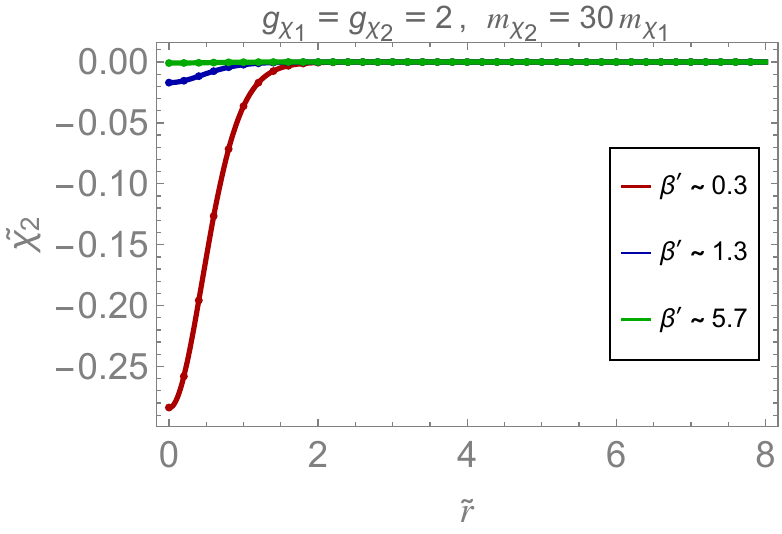}
    \caption{First two rows: Representative plots of fields $\tilde{f}$, $\tilde{\phi}_N$, $\tilde{\chi}_1$ and $\tilde{\chi}_2$ evaluated at $\beta'=0.287\sim0.3$.    
      In one curve, we have the case of only a single mediator $\chi_1$. In the other two curves, we have the case of two mediators $\chi_1$ and $\chi_2$, with $m_{\chi_2}=3\,m_{\chi_1}$ and $m_{\chi_2}=30\, m_{\chi_1}$.
      Final row: Representative plot of $\tilde{\chi}_2$ for different $\beta'$ with $m_{\chi_2}=30\,m_{\chi_1}$. 
      We have set the couplings to be $g_{\chi_1}=g_{\chi_2}=2$.
    }
    \label{fig:compareBeta0.3}
\end{figure}

We have numerically solved this system of equations. A representative plot of the fields is given in Figure \ref{fig:compareBeta0.3}. Here we still define $\beta'$ through the mass of the first field, i.e., $\beta'=m_{\chi_1}\beta$. In the first two rows, we have taken $\beta'\sim 0.3$ for this plot. 
We have set the mass of the second field to be  larger than the mass of the first field; a curve with $m_{\chi_2}/m_{\chi_1}=3$
and a curve with $m_{\chi_2}/m_{\chi_1}=30$.
Also we define $g_{\chi_1}=C_1^2/(4\pi G)$ and $g_{\chi_2}=C_2^2/(4\pi G)$. 
In the last row, we consider different values of $\beta'$ with $m_{\chi_2}=30\,m_{\chi_1}$.
In these plots we have taken $g_{\chi_1}=g_{\chi_2}=2$.


\begin{figure}
    \centering   \includegraphics[width=0.49\linewidth,height=6.2cm]{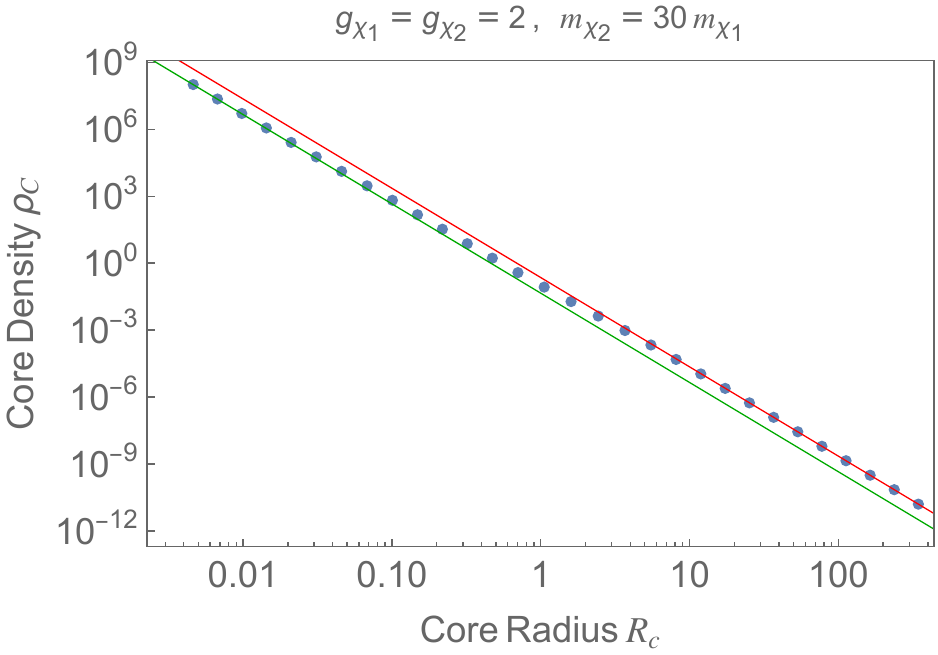}\\
    \vspace{0.2cm}
      \includegraphics[width=0.49\linewidth,height=6.2cm]{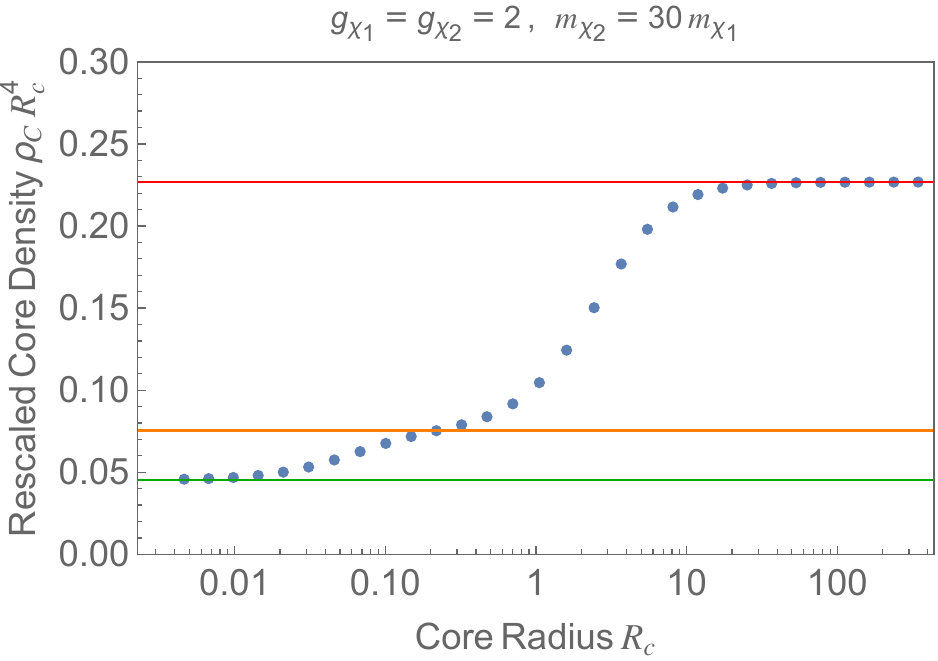}
    \caption{Top: Core density $\rho_c$ (in units of $m_{\chi_1}^4/(G \,m_\phi^2)$) of the soliton versus core radius $R_c$ (in units of $m_{\chi_1}^{-1}$) in the presence of two mediators $\chi_1$ and $\chi_2$. 
    Bottom: Rescaled core density $\rho_c\, R_c^4$ (in units of $(G\, m_\phi^2)^{-1}$) of the soliton versus core radius $R_c$ (in units of $m_\chi^{-1}$).
    Small $R_c\ll m_{\chi_2}^{-1}$, intermediate $m_{\chi_2}^{-1}\ll R_c\ll m_{\chi_1}^{-1}$ and large $R_c\gg m_{\chi_1}^{-1}$ asymptotes are given in green, orange and red lines, respectively.
     We have  set the couplings to be $g_{\chi_1}=g_{\chi_2}=2$
    and the ratio of the masses to be $m_{\chi_2}=30\, m_{\chi_1}$.} 
    \label{fig:2species-Plots}
\end{figure}

The  density versus radius and rescaled density versus radius is given in Figure \ref{fig:2species-Plots}. For a mass hierarchy $m_{\chi_2}\gg m_{\chi_1}$, there are now 3 regimes: 
(i) $R_c\ll m_{\chi_2}^{-1}$, in which both fields are active and the effective strength of gravity is
\beq
G_{\rm eff}=(1+g_{\chi_1}+g_{\chi_2})\,G,\,\,\,\,R_c\ll m_{\chi_2}^{-1}\,.
\eeq
(ii) $m_{\chi_2}^{-1}\ll R_c\ll m_{\chi_1}^{-1}$, in which only field $\chi_1$ is active and the effective strength of gravity is
\beq
G_{\rm eff}=(1+g_{\chi_1})\,G,\,\,\,\,m_{\chi_2}^{-1}\ll R_c\ll m_{\chi_1}^{-1}\,.
\eeq
(iii) $R_c\gg m_{\chi_1}^{-1}$, in which neither field is active and the effective strength of gravity is
\beq
G_{\rm eff}=G,\,\,\,\,R_c\gg m_{\chi_1}^{-1}\,.
\eeq

Each of these regimes is indicated by a corresponding horizontal line: green for (i), orange for (ii), and red for (iii).
In this plot, the ratio of masses is somewhat large, $m_{\chi_2}/m_{\chi_1}=30$, but not very large. One can notice a moderate ``bump" in the shape to the intermediate regime; though this intermediate regime is not very distinct for this moderate mass hierarchy.

One could imagine extending this analysis to many mediators to see a continual rise in the $\rho_c\,R_c^4$ curve as one passes each mass threshold. This could be useful in obtaining a core density versus core radius relation that does not fall so steep, and is moderately more in line with observations. But a detailed comparison to data is beyond the scope of this work.

\section{Discussion}\label{discussion}

In this work we have taken a first step towards understanding the effects of a new force on light scalar dark matter. We have focused on the properties of condensates as a first interesting consequence. We have shown that it alters the core density versus core radius relation. In particular, we find that while we still have the asymptotic relation $\rho_c\propto 1/R_c^4$, the constant of proportionality is different in the $R_c\ll m_\chi^{-1}$ versus $R_c\gg m_\chi^{-1}$ limits. The cross over between these two regimes, softens the slope. This is seen in the rising behavior in the $\rho_c\,R_c^4$ plots given at the bottom of Figure \ref{fig:rhoRrelation}.
In the case of multiple mediators, this rising can be prolonged, as it occurs each time a mass threshold of the mediator is passed, as seen in the bottom of Figure \ref{fig:2species-Plots}.
This can make the theory moderately better in fitting to data, though the improvement is not large here. In particular, we do not  have evidence of an alteration all the way from $q=4$ to $q\approx 1.3$ (with $\rho_c\propto 1/R^q$).
Of interest would be the more dramatic case of significantly larger values of $g_{\chi}$.
We leave a detailed analysis of this scenario and comparison to data for possible future work.
Importantly, this would mean a dramatic departure from the equivalence principle for dark matter on scales less than the Compton wavelength of the mediator $m_\chi^{-1}$. The consequences and  bounds are interesting to explore.

Here we considered the concrete case in which the couplings
$c_1$ and $c_2$ are related by $c_1=c_2/2$. We mentioned that in this special case, the oscillations in the $\phi$ condensate do not produce significant $\chi$ waves, as the leading order coupling is to the energy density $\rho_\phi$ of the condensate which is static in the 
nonrelativistic limit. Future work is to relax this and consider $c_1\neq c_2/2$. A concrete example is $c_1=-c_2/4$, which makes $\chi$ couple to the trace of the energy-momentum tensor of $\phi$, which oscillates as the pressure inside the condensate oscillates with frequency $\approx 2m_\phi$. This can produce $\chi$ waves, whose amplitude can be explored. We note that since the couplings are assumed to be small (of order gravitational strength), this effect may be small too. 

In the presence of this new force, other interesting effects can happen on larger scales too. While the force dies off exponentially on scales $L\gg m_\chi^{-1}$, there may be some residual effects. If $m_\chi^{-1}\sim$\,kpc, then we have seen that the new force can lead to an alteration in the condensate on the scale of the cores of galaxies. This is a scale much smaller than the present Hubble scale. However, one can consider early time effects, where the relevant cosmological scales were much smaller. In particular, at the time of CMB ($t\approx380,000$ years), the Hubble scale was $d_H\approx3t/2\approx 180$\,kpc. On this scale the new force is exponentially suppressed. However, on a scale that is around $\lesssim 1\%$ of the horizon, the new force between dark matter particles would have been appreciable. This is an interesting topic for future work. A possible tool to investigate this can be 
 within the framework of the effective field theory of large scale structure (EFTofLSS) \cite{Baumann:2010tm,Carrasco:2012cv,Hertzberg:2012qn}, where such effects could be systematically parameterized and accounted for.
 Other interesting phenomenology associated with new forces between dark matter can be left for possible future work too.



\section*{Acknowledgments}
M.~P.~H. is supported in part by National Science Foundation grant PHY-2310572.
M.~U. is supported by IBS under the project code, IBS-R018-D3.

\bibliographystyle{IEEEtran}
\bibliography{ref} 

\end{document}